\documentclass{aip-cp}

\usepackage[numbers]{natbib}
\usepackage{graphicx}
\usepackage{epstopdf}

\begin{document}
 
\title{HESS J1826$-$130: A Very Hard $\gamma$-Ray Spectrum Source in the Galactic Plane}

\author[aff1]{E.O.~Ang\"uner\corref{cor1}}
\author[aff2,aff3]{F.~Aharonian}
\author[aff2]{P.~Bordas}
\author[aff1,aff2]{S.~Casanova}
\author[aff4]{C.~Hoischen}
\author[aff5]{I.~Oya}
\author[aff6]{A.~Ziegler}
\author[]{for the H.E.S.S. Collaboration}

\affil[aff1]{Instytut Fizyki J\c{a}drowej PAN, ul. Radzikowskiego 152, 31-342 Krak{\'o}w, Poland}
\affil[aff2]{Max-Planck-Institut f\"ur Kernphysik, P.O. Box 103980, D 69029 Heidelberg, Germany}
\affil[aff3]{Dublin Institute for Advanced Studies, 31 Fitzwilliam Place, Dublin 2, Ireland}
\affil[aff4]{Institut f\"ur Physik und Astronomie, Universit\"at Potsdam,  Karl-Liebknecht-Strasse 24/25, D 14476 Potsdam, Germany}
\affil[aff5]{DESY, Platanenallee 6, D-15738 Zeuthen, Germany}
\affil[aff6]{Friedrich-Alexander-Universit\"at Erlangen-N\"urnberg, Erlangen Centre for Astroparticle Physics, Erwin-Rommel-Str. 1, D 91058 Erlangen, Germany}

\corresp[cor1]{Corresponding author: oguzhan.anguner@ifj.edu.pl}

\maketitle

\begin{abstract}
HESS\,J1826$-$130 is an unidentified hard spectrum source discovered by H.E.S.S. along the Galactic plane, the spectral index being $\Gamma$ = 1.6 with an exponential
cut-off at about 12 TeV. While the source does not have a clear counterpart at longer wavelengths, the very hard spectrum emission at TeV energies implies
that electrons or protons accelerated up to several hundreds of TeV are responsible for the emission. In the hadronic case, the VHE emission can be
produced by runaway cosmic-rays colliding with the dense molecular clouds spatially coincident with the H.E.S.S. source.
\end{abstract}

\section{INTRODUCTION}

In the past decade, observations with the High Energy Stereoscopic System (H.E.S.S.) have discovered tens of very high energy (VHE, $E>0.1$ TeV) $\gamma$-ray 
sources\footnote{See http://tevcat.uchicago.edu for an updated list of known VHE $\gamma$-ray sources.} by scanning a large fraction of the
Galactic Plane \citep{gps}. 

HESS\,J1826$-$130 is an unidentified extended source discovered by the H.E.S.S. telescope array in the multi-TeV domain. The source had been previously hidden in 
the extended tail of emission from the bright nearby source HESS\,J1825$-$137 \citep{J1825}. For such sources affected by source confusion, an investigation in energy bands can
provide an additional powerful tool for new discoveries \citep{igor}. HESS\,J1826$-$130 shows one of the hardest spectra ever observed in VHE $\gamma$-rays, extending up to at 
least few tens of TeV. The shell-type supernova remnants (SNRs), G018.1$-$00.1 and G018.6$-$00.2 \citep{snrs}, are located in the vicinity of the source along with the 
$\gamma$-ray pulsar PSR J1826$-$1256.  

In this work, the analysis of 206 h of H.E.S.S. data from the region around HESS J1826$-$130 will be presented. Finally, possible VHE $\gamma$-ray emission 
scenarios related to the origin of the source will be briefly discussed.

\section{H.E.S.S. OBSERVATIONS and RESULTS}\label{hess}

\subsection{The H.E.S.S. Telescopes}

H.E.S.S. is an array of five imaging atmospheric Cherenkov telescopes located in the Khomas Highland of Namibia, 1800 m above sea level. H.E.S.S. in phase I comprised of four 13 m diameter 
telescopes which have been fully operational since 2004. A fifth telescope was added in the center of the array and has been operational since September 2012. The H.E.S.S. phase I array 
configuration is sensitive to $\gamma$-ray energies between 100 GeV and several tens of TeV. With the addition of the fifth telescope, the energy threshold was lowered to some tens of 
GeV. The VHE H.E.S.S. data presented in this paper were taken with the H.E.S.S. phase I array configuration, which can measure extensive air showers with an angular resolution better than 
${0.1}^{\circ}$ and an average energy resolution of 15$\%$ at an energy of 1 TeV \citep{Aharonian06}. 

\subsection{Detection and Morphological Analysis}\label{morph}

\begin{figure}[h]\label{excess}
  \centerline{\includegraphics[width=300pt]{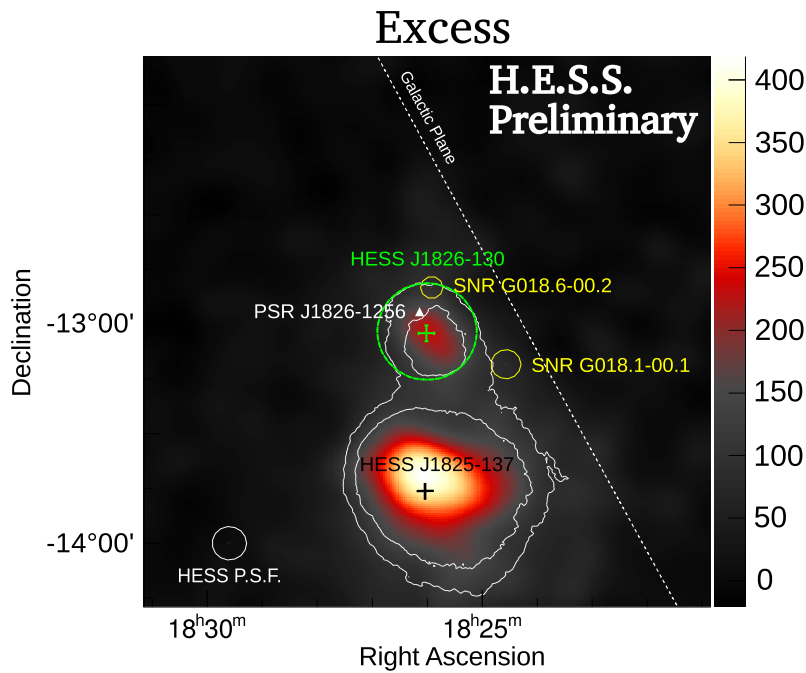}}
  \caption{Map of excess events with energies E $>$ 2 TeV for the region around HESS\,J1826$-$130 smoothed with the H.E.S.S. PSF. The white contours indicate the significance of the emission 
at the 15$\sigma$ and 20$\sigma$ level. The color scale is in units of excess counts per smoothing Gaussian width. The 68$\%$ containment radius of the H.E.S.S. PSF is shown with the white 
circle at the lower left corner. The green circle shows the integration region used for deriving the source spectrum shown in Fig. \ref{sp}, while the green cross indicates the value and 
1$\sigma$ uncertainty of the best fit position of the source. The nearby SNRs, G018.6$-$00.2 and G018.1$-$00.1 are marked with yellow circles, while the white triangle indicates the position 
of the $\gamma$-ray pulsar PSR J1826$-$1256.}
\end{figure}

The observations of the field of view around HESS\,J1826$-$130 were carried out between 2004 and 2015, corresponding to an acceptance corrected live-time of 206 h of H.E.S.S. phase I data 
after the application of the quality selection criteria \citep{Aharonian06}. The data have been analyzed with the H.E.S.S. analysis package for shower reconstruction and the multivariate 
analysis technique \citep{TMVA} has been applied for providing an improved discrimination between hadrons and $\gamma$ rays. In order to provide improved angular resolution and reduce 
contamination from the bright and nearby source HESS\,J1825$-$137, the source position and morphology have been obtained with a hard cut configuration which requires a minimum 
of 160 photo-electrons per image. The cosmic-ray background level was estimated using the ring background model \citep{berge2007}. Using this dataset, HESS\,J1826$-$130 is detected with a 
statistical significance of 21.0$\sigma$, determined by using Equation (17) in \cite{lima}. Figure \ref{excess} shows the acceptance corrected and smoothed with the H.E.S.S. point spread 
function (PSF) VHE $\gamma$-ray excess map of the region around HESS\,J1826$-$130 at energies greater than E = 2 TeV.

\begin{figure}[ht]\label{engDep}
  \centerline{\includegraphics[width=300pt]{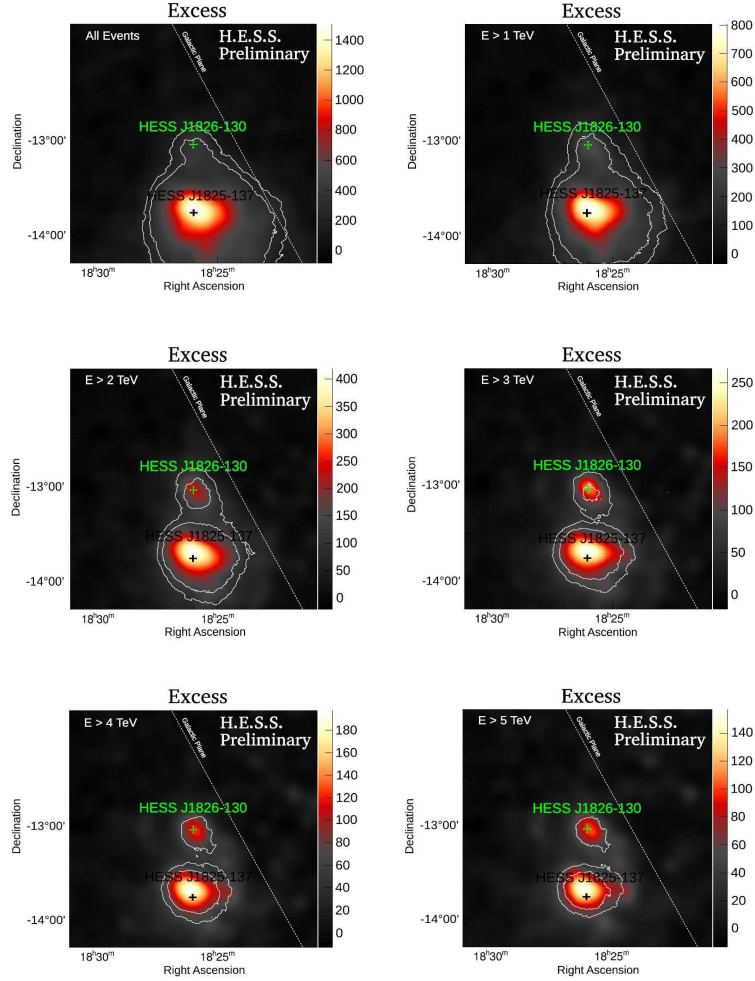}}
  \caption{Energy dependent VHE $\gamma$-ray excess maps of the region around HESS\,J1826$-$130 for the same field of view given in Fig. \ref{excess}.}
\end{figure}

The source remained unnoticed and hidden for the standard source detection techniques due to its relatively low brightness with respect to the nearby bright source HESS\,J1825$-$137. During a 
study of a possible energy-dependent morphology of HESS\,J1826$-$130, a collection of images for events with energies above a set of energy thresholds (E\,$>$\,1, 2, 3, 4, and 5 TeV) was 
created (see Fig. \ref{engDep}). As it can be seen from the figure, HESS\,J1826$-$130 becomes more visible with increasing energy threshold. 

The position and extension of HESS\,J1826$-$130 was determined by fitting a two-dimensional Gaussian model convolved with the H.E.S.S. PSF to the excess event distribution. The centroid of 
the 2D Gaussian corresponding to the best fit position of the source is R.A. (J2000): $18^{\textsuperscript{h}}$26$^{\textsuperscript{m}}$0.2$^{\textsuperscript{s}}$ $\pm$ 
7.0$^{\textsuperscript{s}}_{\rm stat}$ $\pm$ 1.3$^{\textsuperscript{s}}_{\rm sys}$ and Dec. (J2000): $-$13$^{\circ}$02' $\pm$ 1.8'$_{\rm stat}$ $\pm$ 0.3'$_{\rm sys}$. The extension of the 
source is estimated as 0.17$^\circ$ $\pm$ 0.02$^{\circ}_{\rm stat}$ $\pm$ 0.05$^{\circ}_{\rm sys}$.    

\subsection{Spectral Analysis} \label{spect}

A circular region with a radius of 0.22$^{\circ}$ centered at the best fit position of HESS J1826$-$130 (e.g. Fig. \ref{excess}, green circle) was used as an integration region for extracting 
the differential VHE $\gamma$-ray spectrum of the source. The spectrum was derived using the forward folding technique \citep{piron2001}, and is well described by a power-law with an 
exponential cut-off function $dN/dE$ = $\Phi_{0}$ ($E/1\textnormal{ TeV}$)$^{-\Gamma}$ $\exp$($-E/E_{c}$) with $\Phi_{0}$ = (8.62 $\pm$ 0.73$_{\rm stat}$ $\pm$ 1.72$_{\rm sys}$) 
$\times$ $10^{13}$ cm$^{-2}$ s$^{-1}$ TeV$^{-1}$, $\Gamma$ = 1.61 $\pm$ 0.11$_{\rm stat}$ $\pm$ 0.20$_{\rm sys}$ and a cut-off energy of $E_{c}$ = 12.5$^{+4.0}_{-2.4}$ TeV. This spectral 
model is preferred over a simple power-law model at a statistical significance level of 4.0$\sigma$ from log-likelihood ratio test. The time variability of the emission of the source was 
investigated at different timescales by fitting the integral flux to a constant. In all the timescales, from run-wise ($\sim$30 min) to year-wise light curves, the $\chi^{2}$ fit results in 
compatibility of the emission with a constant flux. 

\begin{figure}[ht]\label{sp}
  \centerline{\includegraphics[width=400pt]{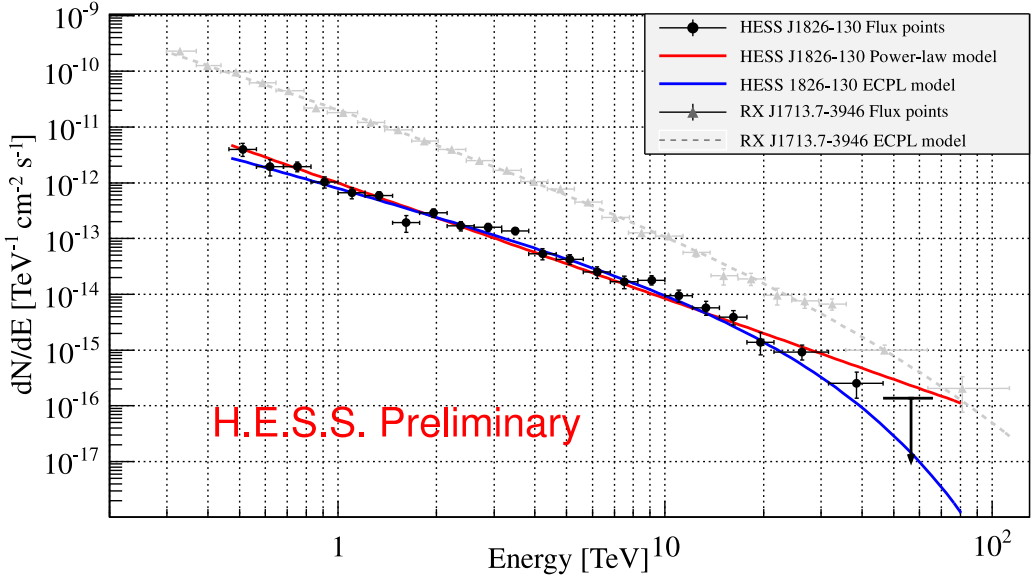}}
  \caption{VHE $\gamma$-ray spectrum of HESS\,J1826$-$130 extracted from the source region shown in Fig. \ref{excess}. The black dots show the flux points with errors. The 68$\%$ error band of 
the best fit ECPL model (blue shaded region) and the best fit power-law model (red line) are also shown. The differential flux upper limit in the shown energy band is at 99$\%$ confidence 
level. The VHE spectrum of RXJ1713.7$-$3946 is shown for comparison.}
\end{figure}

The integration region used for extracting the VHE spectrum of HESS\,J1826$-$130 is strongly contaminated by HESS\,J1825$-$137, especially at lower energies ($\sim$40$\%$ below 1.5 TeV and 
$\sim$20$\%$ above 1.5 TeV), and the spectrum presented in this section is affected by this contamination. Further spectral studies (simultaneous fitting of both source spectra) are being 
performed for obtaining the intrinsic spectrum of HESS\,J1826$-$130. Given the relatively softer spectrum of HESS\,J1825$-$137, the intrinsic spectrum of HESS\,J1826$-$130 is expected to be 
even harder with respect to the spectral results presented in this section.   

\section{MULTIWAVELENGTH COUNTERPARTS}\label{mwl}

Two giant molecular clouds (GMC) with masses of few 10$^{5}$ M$_{\bigodot}$ at 3.7 kpc and 4.7 kpc have been found along the line of sight of to HESS\,J1826$-$130 when studying 
NANTEN data \citep{voisin}. The ambient gas densities of these GMCs are estimated of the order of $\sim$600 cm$^{-3}$. 

The SNR G018.1$-$00.1 is located between 4 and 6.3 kpc \citep{snrs} (corresponding to the projected physical distance of $\sim$33 pc from the best fit position of HESS\,J1826$-$130), while 
the SNR G018.6$-$00.2 is located between 4 and 5.2 kpc \citep{snrs} (corresponding to the projected physical distance of $\sim$18 pc from the best fit position of HESS\,J1826$-$130).

The $\gamma$-ray pulsar PSR\,J1826$-$1256 is also located in the vicinity of the source as shown in Fig. \ref{excess}. This pulsar has a spin-down luminosity of $\dot{\textnormal{E}}$ = 3.6 $\times$ 
10$^{36}$ erg/s and a characteristic age of $\tau_{c}$ = 14.4 kyr. Note that pulsars with $\tau_{c}$ $<$ 150 kyr and with $\dot{E}$/$D_{\rm PSR}^{2}$ $\geq$ 10$^{34}$ erg 
s$^{-1}$ kpc$^{-2}$ are known to power pulsar wind nebulae (PWNe) that are detectable at very high energies \citep{pwnpopB}. This pulsar has been detected both in X-ray energies \citep{xray} 
and in high energy $\gamma$-rays \citep{he}. No distance estimation for this pulsar is known from dispersion measure \citep{atnf} since it is a radio-quiet pulsar. 

\section{DISCUSSION}\label{dis}

The emission from HESS\,J1826$-$130 is characterized by a very hard spectrum ($\Gamma$ = 1.6) with an exponential cut-off at about 12 TeV. Such a spectrum, together with a spatial coincidence
of a dense molecular hydrogen region, suggests that the radiation could be produced by several hundred TeV protons colliding with the gas. To explain the very hard spectrum, one
can think of a hadronic scenario in which runaway protons accelerated by the close shell-type SNRs G018.1$-$00.1 or G018.6$-$00.2 are emitting TeV photons when interacting with the dense 
ambient gas found along the line of sight. The proton spectrum should extend up to at least several hundreds of TeV. The $\gamma$-ray luminosity, L$_{\gamma}$, of the source is 8
 $\times$ 10$^{33}$ ergs for a distance of 4 kpc. This translates into an energy output in accelerated protons, W$_{\textnormal{pp}}$ = L$_{\gamma}$ $\times$ t$_{\textnormal{pp}}$, of 
6 $\times$ 10$^{49}$ (n / 1 cm$^{-3}$)$^{-1}$ erg, while ambient gas densities from NANTEN data analysis are of the order of 600 cm$^{-3}$. In the hadronic scenario, the hardness of the photon 
spectrum could be the effect of the highest energy protons diffusing in dense clouds, while the lower energy protons might be still confined within the accelerating source or efficiently 
excluded from the clouds if diffusion coefficient inside the clouds is suppressed.

A leptonic scenario, where electrons accelerated by the pulsar PSR\,J1826$-$1256 are up-scattering CMB or IR photons, can also explain the VHE emission. Such as hard spectrum at H.E.S.S. 
energies can be produced by an uncooled electron population with spectral index close to -2.0 and a cutoff at around 70 TeV. The energy output (W$_{\textnormal{e}}$) in accelerated electrons 
is 2 $\times$ 10$^{47}$ ergs for a distance of 7 kpc. This source has a spectrum very similar to other few PWNe, in particular, Vela X \citep{vela}. HESS\,J1826$-$130 could be an indication 
of a distinctive PWN population, with very hard spectra and relatively high cutoff energies

\section{CONCLUSIONS}\label{conc}

The unidentified TeV source, HESS\,J1826$-$130, stands out in the H.E.S.S. Galactic Plane Survey for its particularly hard spectrum. A proton or electron population extending up to several 
hundred TeV is needed to produce the TeV emission from HESS\,J1826$-$130. Currently, studies to better constrain its spectrum are under way.

\section{ACKNOWLEDGMENTS}
The support of the Namibian authorities and of the University of Namibia in facilitating the construction and operation of H.E.S.S. is gratefully acknowledged, as is the support by the German 
Ministry for Education and Research (BMBF), the Max Planck Society, the German Research Foundation (DFG), the French Ministry for Research, the CNRS-IN2P3 and the Astroparticle 
Interdisciplinary Programme of the CNRS, the U.K. Science and Technology Facilities Council (STFC), the IPNP of the Charles University, the Czech Science Foundation, the Polish Ministry of 
Science and Higher Education, the South African Department of Science and Technology and National Research Foundation, the University of Namibia, the Innsbruck University, the Austrian Science 
Fund (FWF), and the Austrian Federal Ministry for Science, Research and Economy, and by the University of Adelaide and the Australian Research Council. We appreciate the excellent work of the 
technical support staff in Berlin, Durham, Hamburg, Heidelberg, Palaiseau, Paris, Saclay, and in Namibia in the construction and operation of the equipment. This work benefited from services 
provided by the H.E.S.S. Virtual Organisation, supported by the national resource providers of the EGI Federation. This research has made use of software provided by the {\it Chandra} X-ray 
Center (CXC) in the application packages CIAO, ChIPS, and Sherpa. This research has made use of the SIMBAD database, operated at CDS, Strasbourg, France. This research has made use of the 
ATNF pulsar catalog database (http://www.atnf.csiro.au/research/pulsar/psrcat/). The NANTEN project is based on the mutual agreement between Nagoya University and the Carnegie Institution of 
Washington. Sabrina Casanova and Ekrem O\u{g}uzhan Ang\"uner acknowledge the support from the Polish National Science Center under the Opus Grant UMO-2014/13/B/ST9/00945.

\end{document}